\documentclass{PoS}

\usepackage{array,graphicx}

\newcommand{\Bd }   {\ensuremath{B^0}}

\newcommand{\Bs   } {\ensuremath{B^0_s}}
\newcommand{\KK   } {\ensuremath{K^+K^-}}
\newcommand{\bBs  }{\ensuremath{\bar{B}^0_s}}

\newcommand{\BsMix}{\Bs-\bBs}
\newcommand{\Jpsi}{\ensuremath{J\!/\!\psi}}
\newcommand{\fnot} {\ensuremath{f_0(980)}}
\newcommand{\Kstar}{\ensuremath{K^{\star}\!(892)^{0}}}

\newcommand{\Kpi}{\ensuremath{K^+\pi^-}}

\newcommand{\myto}{\kern -0.3em\to\kern -0.2em}
\def\BsJpsiPhi{\mbox{\ensuremath{\Bs \myto \Jpsi\,\kern -0.1em\phi}}}
\newcommand{\BsPhiPhi}{\ensuremath{\Bs \myto \phi \,\kern -0.1em\phi}}
\newcommand{\BsJpsiKK}{\ensuremath{\Bs \myto \Jpsi \,\kern -0.1em\KK}}
\newcommand{\BsJpsifnot}{\ensuremath{\Bs \myto \Jpsi \,\kern -0.1emf_0}}
\newcommand{\BdJpsiKpi}{\ensuremath{\Bd \myto \Jpsi \,\kern -0.1em\Kpi}}
\newcommand{\BdJpsiKstar}{\ensuremath{\Bd \myto \Jpsi\,\kern -0.1em \Kstar}}
\newcommand{\BsDsDs}{\ensuremath{\Bs \myto D^{(\star)+}_s D^{(\star)-}_s}}

\newcommand{\phis    } {\ensuremath{\phi_s}}
\newcommand{\DGs     }  {\ensuremath{\Delta\Gamma_s}}

\newcommand{\CP  } {\ensuremath{C\!P}}   
\newcommand{\ACP}{\ensuremath{\mathcal{A}_{\CP}}}

\newcommand{\eg}{\textit{e.\,g.}}
\newcommand{\stat}{\ensuremath{{\rm (stat)}}}
\newcommand{\syst}{\ensuremath{{\rm (syst)}}}
\newcommand{\br}{\ensuremath{{\rm (BR)}}}

\newcommand{\tevatron}{Tevatron}

\newcommand{\dEdx}{\ensuremath{dE\!/\!dx}}

\newcommand{\babar}{\mbox{\slshape B\kern-0.1em{\smaller A}\kern-0.1em B\kern-0.1em{\smaller A\kern-0.2em R}}}

\newcommand{\degrees}{\ensuremath{^\circ}}

\newcommand{\tev}{\ensuremath{\mathrm{Te\kern -0.1em V}}}
\newcommand{\gev}{\ensuremath{\mathrm{Ge\kern -0.1em V}}}
\newcommand{\mev}{\ensuremath{\mathrm{Me\kern -0.1em V}}}
\newcommand{\kev}{\ensuremath{\mathrm{ke\kern -0.1em V}}}
\newcommand{\massgev}{\mbox{\gev/$c^2$}}
\newcommand{\massmev}{\mbox{\mev/$c^2$}}
\newcommand{\pgev}{\mbox{\gev/$c$}}

\newcommand{\lumifb}{\mbox{fb$^{-1}$}}				%	fb^-1

\newcommand{\Journal}[4]{{#1} {\bf #2}, #3 (#4)}

\newcommand{\PLB}{{\em Phys. Lett.}  B}
\newcommand{\PRL}{\em Phys. Rev. Lett.}
\newcommand{\PRD}{{\em Phys. Rev.} D}

\title{\CP\ Violation in Hadronic $B$ Decays at CDF}

\ShortTitle{\CP\ Violation in Hadronic $B$ Decays at CDF}

\author{Mirco Dorigo\thanks{Speaker on behalf of the CDF collaboration.}\\
       INFN and University of  Trieste\\
       E-mail: \email{mirco.dorigo@ts.infn.it}}

\abstract{I report  recent measurements in $b$-hadron decays reconstructed 
in the full data set of $\sqrt{s} = 1.96$\,\tev\  proton-antiproton collisions
collected by the CDF experiment at the Tevatron. These include the final CDF results on: 
 measurements of  \CP\ asymmetries in two-body charmless decays
of the \Bd, \Bs, and $\Lambda^0_b$ hadrons; bounds on the \Bs\ mixing phase 
and on the decay width difference of \Bs\ mass eigenstates; 
and updated measurements of branching ratios of \BsJpsiPhi\ and \BsDsDs\ decays.
All measurements are among the most precise from a single experiment and in 
agreement with the standard model predictions.
}

\FullConference{36th International Conference on High Energy Physics,\\
		July 4-11, 2012\\
		Melbourne, Australia}

\begin{document}

\section{Direct \CP-violation in charmless two-body $b$-hadron decays}
Recently, the pattern of direct \CP\ violation in
charmless mesonic decays of $B$ mesons has shown some 
unanticipated discrepancies from expectations that could
be accommodated in several simple extensions of the standard model (SM) \cite{Bhh1}. 
However, uncertainties on the contribution of higher-order SM
amplitudes has prevented a firm conclusion.
Redundant and precise measurements in similar decays related by flavor symmetries provide
additional constraints to reduce the theoretical uncertainties 
and probe non-SM physics in the electro-weak amplitudes.
%High-precision measurements of \CP\ violation in
%such decay modes remains therefore very interesting.
Specifically, measurements of direct \CP\ violation in $\Bs \myto K^-\pi^+$
decays have been proposed as a nearly model-independent test for the presence of non-SM physics \cite{Bhh2}. 
The Cabibbo-Kobayashi-Maskawa (CKM) mechanism 
predicts a well-defined hierarchy between direct \CP\ violation in $\Bd \myto K^+\pi^-$
and $\Bs \myto K^-\pi^+$ decays, yielding a significant asymmetry for the latter, of about 30\%. 
Supplementary information could come from \CP\ violation in bottom baryons:  
\CP-violating asymmetries in charmless $\Lambda_b^0$ decays could contribute additional insight.
 
The displaced vertex trigger of CDF is effective
at selecting two-body $b$-hadron decays \cite{SVT}.
We report a recent measurement of \CP\ asymmetries of
such decays which uses all of the data from
Run II of the \tevatron, corresponding to about 9.3\,\lumifb\
of integrated luminosity \cite{Bhh3}. 
The offline selection is an unbiased and optimized set of requirements inherited from an earlier analysis \cite{Bhh4}, 
which relies on a more accurate determination of the same quantities used in the trigger
(\eg, two opposite-charged tracks with impact parameter between 0.1 and 1\,mm; opening angles 
between tracks in the range $20\degrees$ and $135\degrees$; $B$ decay length in the transverse 
plane greater than 0.2\,mm) with the
addition of two further observables: the isolation of the $B$ candidate and the quality of the three-dimensional
fit of the decay point of the $B$ candidate.  
 No more than one $B$ candidate per event is 
found after the selection, and a mass is assigned to each, using
a charged pion mass assignment for both decay products.
The resulting $\pi^+\pi^-$ spectrum is shown in Fig.~\ref{fig:results}~(left).
Backgrounds include mis-reconstructed multibody $b$-hadron decays, causing 
the shoulder at 5.16\,\massgev, and random pairs of charged particles. 
In spite of the $25$\,\massmev\ mass resolution of CDF detector for two-body hadronic decays,
the various signal modes overlap into an unresolved mass peak near the nominal \Bd\ mass, 
with a width of about $35$\,\massmev.

We use an unbinned likelihood fit, incorporating kinematic 
and particle-identification (PID) information, to
determine the fraction of each individual signal channel and the charge asymmetries, 
uncorrected for instrumental effects, of the flavor-specific decays $\Bd \myto K^+\pi^-$, 
$\Bs \myto K^-\pi^+$, $\Lambda_b^0 \myto p\pi^-$, and $\Lambda_b^0 \myto p K^-$.
The decay flavor is inferred from the charges of final state particles assuming 
equal numbers of $b$ and $\bar{b}$ quarks at production. 
Any effect from \CP\ violation in $B$-meson flavor mixing is assumed to be negligible.
The likelihood exploits the kinematic differences among the decay modes,  
enhancing statistical separation power between $\pi\pi$ and $KK$ (or $K\pi$) final states.
The kinematic information is summarized by three loosely correlated variables: 
the square of the invariant mass $m^2_{\pi\pi}$; 
the signed momentum imbalance $\beta = (p^+ - p^-)/(p^+ + p^-)$, 
where $p^+$ ($p^-$) is the the momentum of the positive
(negative) particle; and the scalar sum of particle momenta $p = p^+ + p^-$.
Kinematic fit templates are extracted from
simulation for signal and physics background, while they are extracted from an independent 
sample for combinatorial background. 
Mass line shapes are accurately described for the non-Gaussian resolution tails and for the
effects of the final state radiation of soft photons.
This resolution model was checked against the observed shape
of the 3.8 million $D^0 \myto K^-\pi^+$ decays in a sample of $D^{\star +} \myto D^0\pi^+$
decays collected with a similar trigger selection. This
sample was also used to calibrate the PID  through the measurement 
of the specific ionization energy-loss (\dEdx) of  kaons and pions in the drift chamber,
using the charge of the $D^{\star +} $ pion to identify the $D^0$
decay products. The \dEdx\ response of protons was determined
from a sample of about 0.3 million $\Lambda \myto p\pi^-$
decays, where the kinematics and the momentum threshold of the trigger
allow unambiguous identification of the decay products.  
The statistical separation between kaons and pions
is about 1.4$\sigma$, while the ionization rates of protons and kaons 
are quite similar in the momentum range of interest.
The signal yields from the fit are corrected for different detection efficiencies 
extracted from control samples in data to determine the physical asymmetries. 
Simulation is used only to account for small differences between 
the kinematics of decays and control signals. 
The corrections for decays into the $K^+\pi^-$ final state
are extracted from a sample of about 30 million
untagged $D^0 \myto K^-\pi^+$ decays.
 For the $\Lambda^0_b \myto p \pi^-$
asymmetry, the factor is derived from a control 
sample of $\Lambda \myto p\pi$ decays. 
The $pK^-$ factor is extracted by combining the previous ones and assuming the 
trigger and reconstruction efficiency for two particles factorizes as
the product of the single-particle efficiencies.

The final results are listed in Tab.~\ref{tab:results}. The asymmetry
$\ACP(\Bd\myto K^+\pi^-)$ is consistent with results from $B$-factories
and LHCb. The measured $\ACP(\Bs\myto K^-\pi^+)$
 confirms the LHCb evidence with the same level of resolution.
 Systematic uncertainties on such asymmetries are largely due to
variation of the fit results against different templates for combinatorial background,
signal mass distributions and PID modeling.
The observed asymmetries of the $\Lambda^0_b$ decays are consistent with zero. 
Their systematic uncertainty is dominated by the 
variation of unknown polarization amplitudes in the templates.
They are in agreement with the previous results from CDF and supersede them \cite{Bhh4}.

\section{Measurement of the \BsJpsiPhi\ time-evolution and branching ratio in the complete CDF dataset}
The \BsMix\ mixing is a promising process for searches for new physics (NP), 
given the D0 3.9$\sigma$ anomaly in dimuon charge asymmetry \cite{Asl_d0}. 
 If the anomaly is due to new dynamics in the \Bs\ sector  
the phase difference between the \BsMix\ mixing amplitude
and the amplitude of  \Bs\ and \bBs\ decays into common 
final states, \phis, would be significantly  
altered with respect to its nearly vanishing value expected in the SM. 
A non-CKM enhancement of \phis\ can also decrease 
the decay width difference \DGs\ between the heavy 
and light mass-eigenstates of the \Bs\ meson.
The analysis of the time evolution of \BsJpsiPhi\ decays 
is the most effective experimental probe of such a \CP-violating phase. 
Since the decay is dominated by a single real amplitude, 
the phase difference equals the mixing phase to a good approximation. 
Early \tevatron\ measurements have shown a mild discrepancy 
of about 2$\sigma$ with the SM expectation \cite{phis_cdf_d0_comb}. 
Latest updates by CDF and D0 are in better agreement with the SM, 
as well as measurements provided by LHCb \cite{phis_cdf_5fb,phis_d0_8fb,phis_lhcb,phis_atlas}.

Here we report the new CDF update using the complete dataset of 9.6\,\lumifb\ which comprises
about 11\,000 \BsJpsiPhi\ decays collected by the dimuon trigger \cite{phis_cdf_new}. 
The decays are fully reconstructed through four tracks that fit to a common decay point separated from the beam line, 
two matched to muon pairs consistent with a \Jpsi\ decay, and two consistent with a $\phi \myto\KK$ decay.
  A joint fit that exploits the candidate-specific information given by the 
  $B$ mass, decay time and production flavor, along with the 
 decay angles of kaons and muons, is used to determine both \phis\ and \DGs. 
 The analysis closely follows the previous determination obtained on a subset of the data \cite{phis_cdf_5fb}. 
 The only major difference is the use of an updated calibration of  the 
 tagging algorithm that uses information from the decay of the opposite side bottom hadron
 in the event to determine the flavor of the \Bs\ at its production, with tagging power $(1.39 \pm 0.05)\%$. 
 The information from the tagger that exploits charge-flavor correlations of
 the neighboring kaon to the \Bs\ is instead restricted to only half of the sample, 
 in which has tagging power $(3.5 \pm 1.4)\%$, because its calibration 
 is feasible for early data only. 
 This degrades the statistical resolution on \phis\ by no more than 15\%.
 A decay-time resolution of $\sim$90\,fs allows resolving 
 the fast \Bs\ oscillations to increase sensitivity on the mixing phase. 
  
  Included in the analysis is also the \CP-odd $S$-wave component which  
originates from the non-resonant \KK\ pair or from \fnot\ decays. 
The fraction of $S$-wave in the \KK\ mass range 1.009--1.028\,\massgev\ 
is determined from the angular information to be consistent with zero with 
$\mathcal{O}(2\%)$ uncertainty, which is in agreement with our previous determination \cite{phis_cdf_5fb}
and the LHCb \cite{phis_lhcb} and ATLAS results \cite{phis_atlas}, and inconsistent with 
the D0 determination \cite{phis_d0_8fb}. 
An auxiliary simultaneous fit of the \Jpsi\KK\ and \KK\ mass distributions, 
which includes for the first time the full resonance structure of the \BdJpsiKpi, determines 
a $(0.8 \pm 0.2\stat)\%$ \KK\ $S$-wave contribution, in agreement with the central fit. 
The contamination from mis-identified \Bd\ decays is $(8.0 \pm 0.2\stat)\%$, 
which is significantly larger than the 2\% values derived assuming only $P$-wave \Bd\ decays \cite{phis_cdf_5fb}.  
If neglected, this additional \Bd\ component could mimic a larger \KK\ $S$-wave than is present. 
 
\begin{figure}[t]
\begin{center}
\includegraphics[width=.38\textwidth]{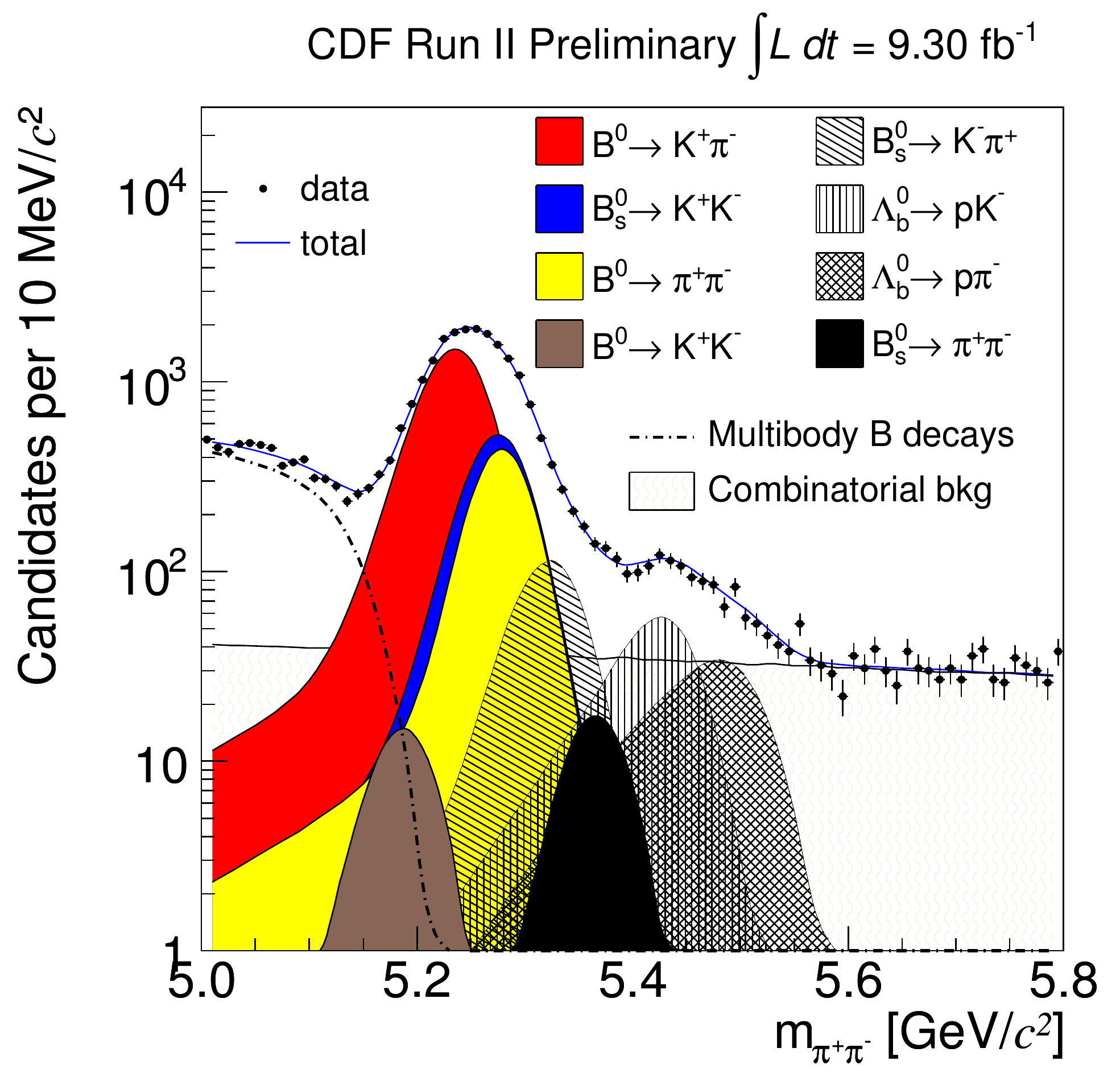}\hfil
 \includegraphics[width=.355\textwidth]{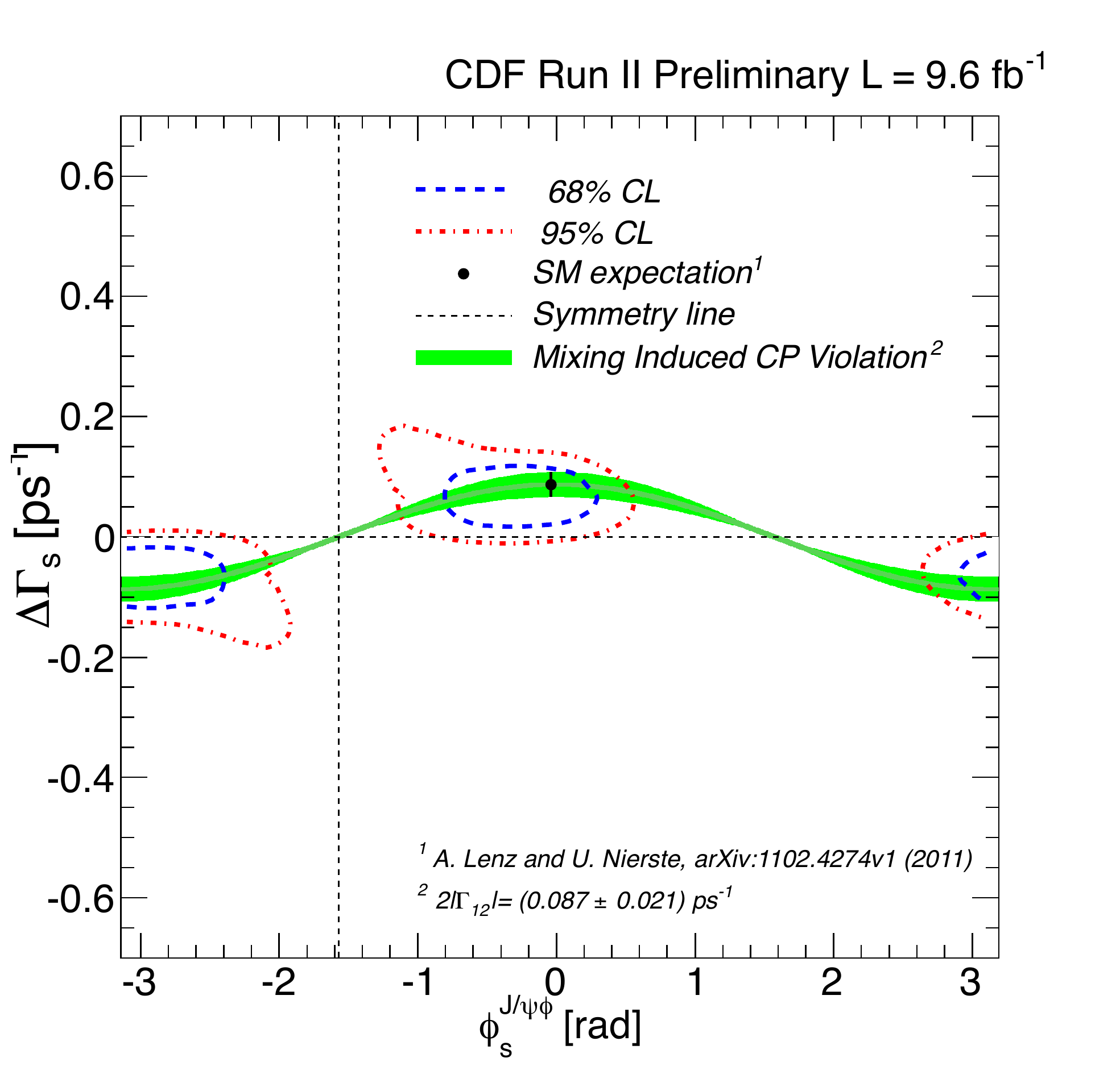}
\end{center}
\caption{Distribution of $\pi^+\pi^-$ spectrum with fit projections overlaid (left); 
68\%  and 95\% confidence regions in the plane (\phis,\DGs) from profile-likelihood of CDF data (right).
 }
\label{fig:results}
\end{figure}

The 68\% and 95\% confidence regions in the plane $(\phis,\DGs)$
 obtained using a profile-likelihood ratio ordering with frequentist inclusion 
 of systematic uncertainties are reported in Fig.~\ref{fig:results}~(right).   
The confidence interval for the mixing phase is
 $\phis \in [-0.60,0.12]$ rad at 68\% C.L.,
 in agreement with the CKM value and other recent determinations \cite{phis_d0_8fb,phis_lhcb,phis_atlas}.
 This is the final CDF measurement on the \Bs\ mixing phase, and provides a factor 40\% improvement 
 in resolution with respect to the latest determination.   
In Tab.~\ref{tab:results}, we also report  the measurements of \DGs\ and of the \Bs\ lifetime $\tau_s$,
 under the hypothesis of a SM value for \phis, which are in agreement 
 with other experiments' results \cite{phis_d0_8fb,phis_lhcb,phis_atlas}.

The Run II dimuon data set is also exploited for measuring 
the ratio of the \BsJpsiPhi\ and \BdJpsiKstar\ branching fractions,
$R=(f_s/f_d)(\mathcal{B}(\BsJpsiPhi)/\mathcal{B}(\BdJpsiKstar))$,
where $f_{s(d)}$ is the fragmentation fraction of the $s(d)$ quark \cite{BRref}. 
We adopt an offline selection that differs
from the selection for the \phis\ measurement since it is optimized towards the measurement of $R$. 
We extract the \Bs\ and \Bd\ yields through a 
fit to the binned \Jpsi\KK\ and \Jpsi\Kpi\ invariant mass distributions. 
After application of a relative acceptance factor determined from simulation, 
we measure $R=0.239 \pm 0.003\stat \pm 0.019\syst$. 
Systematic uncertainties are dominated by the modeling 
of the background and signal templates in the fit and 
by the variation within uncertainties of the world average value of 
$B$ lifetimes and polarization amplitudes in the
simulation for the acceptance's estimation. 
Using the most recent CDF measurement \cite{CDFfsfd} of $f_s/(f_u + f_d)  \mathcal{B}(D_s \myto \phi\pi)$ 
combined with the PDG value of $\mathcal{B}(D_s \myto \phi\pi)$~\cite{pdg} to extract $(f_s/f_d)$ and
the PDG value of $\mathcal{B}(\BdJpsiKstar)$~\cite{pdg}, we find the branching ratio $\mathcal{B}(\BsJpsiPhi)$
reported in Tab.~\ref{tab:results}, in agreement
with and with similar resolution to Belle's latest result \cite{belleBR}. 
Using the latter, we also compute the ratio $f_s/f_d$ (Tab.~\ref{tab:results}).
The analysis is also performed in bins of \Bs\ transverse momentum 
in a range of $6<p_T <27$\,\pgev, showing no dependence of $f_s/f_d$ versus $p_T$.  

\section{Measurement of \BsDsDs\ branching ratios}
A measurement of \Bs\ production rate times 
the \BsDsDs\ branching ratio relative to the normalization 
mode $B^0 \myto D^{+}_s D^{-}$ is perfomed using a data sample 
corresponding to an integrated luminosity of 6.8\,\lumifb\
recorded by the displaced track trigger~\cite{BsDsDsref}.
The decays $D^+_s \myto K^+ K^- \pi^+$ and $D^+ \myto K^- \pi^+ \pi^+$ 
 are reconstructed from combinations of three tracks 
with appropriate charge and mass hypothesis assignments, 
fitted to a common vertex and then combined into another vertex to 
form \Bs\ candidates. For the first time in this channel, the Dalitz structure of the intermediate 
states is exploited for an accurate evaluation of acceptances and efficiencies.
The relative branching fractions are determined in a simultaneous 
maximum likelihood fit to mass distributions of two signal samples, $(\phi \pi^+)(\phi \pi^-)$ and 
$(\bar{K}^{\star}\!(892)^0 K^+)(\phi \pi^-)$, and two normalization samples, $(\phi \pi^+)(K^+ \pi^- \pi^-)$ 
and $(\bar{K}^{\star}\!(892)^0 K^+)(K^+ \pi^- \pi^-)$. Using measured values of production and relative branching fractions, 
the absolute branching fractions shown in Tab.~\ref{tab:results} are derived.
These results are the most precise to date from a single experiment and can provide 
information on the decay width difference \DGs .

\begin{table}[t]
\centering   
\newcolumntype{L}{>{$\scriptstyle}l<{$}}
\newcolumntype{C}{>{$\scriptstyle}c<{$}}         
\begin{tabular}{LCc}
	\hline\hline
	\ACP(\Bd\to K^+\pi^-) & (-8.3 \pm 1.3\stat \pm 0.3\syst)\%  \\
	\ACP(\Bs\to K^-\pi^+) & (22 \pm 7\stat \pm 2\syst)\%  \\
	\ACP(\Lambda_b^0\to p\pi^-) & (7 \pm 7\stat\pm 3\syst)\%  \\
	\ACP(\Lambda_b^0\to p K^-) & (9 \pm 8\stat \pm 4\syst)\%  \\
									\hline
	\phis & [-0.60,0.12]\,\,\rm{at\,\,68\%\,\,C.L.}  \\
	\DGs & 0.068 \pm 0.026\stat \pm 0.009\syst\,\,\rm{ps}^{-1} \\
	\tau_s & 1.528 \pm 0.019\stat \pm 0.009 \syst\,\,\rm{ps}\\ 
	\hline
       \mathcal{B}(\Bs\to\Jpsi\phi) & (0.118 \pm 0.002\stat \pm 0.009\syst \pm 0.015\br)\%\\
		f_s/f_d 		& 0.254 \pm 0.003\stat\pm 0.020\syst\pm0.044\br  \\ 	
\hline
\mathcal{B}(\Bs \to D^+_s D^-_s)            & (0.49 \pm 0.06\stat \pm 0.05\syst \pm 0.08\br)\%   \\ 
\mathcal{B}(\Bs \to D^{\star\pm}_s D^{\mp}_s)    & (1.13 \pm 0.12\stat  \pm 0.09\syst \pm 0.19\br)\% \\
\mathcal{B}(\Bs \to D^{\star+}_s D^{\star-}_s)       & (1.75 \pm 0.19\stat  \pm 0.17\syst \pm 0.29\br)\% \\
\mathcal{B}(\Bs \to D^{(\star)+}_s D^{(\star)-}_s)   & (3.38 \pm 0.25\stat  \pm 0.30\syst \pm 0.56\br)\% \\				
\hline\hline				
\end{tabular}
\caption{\label{tab:results} Summary of results of the measurements described in this report. }
\end{table}

\section{Conclusions}
I have reported recent measurements in $b$-hadron decays which exploit the full CDF data set.
The measured \CP\ asymmetry of $\Bs\myto K^-\pi^+$ decays
confirms the LHCb result with similar resolution, providing strong evidence of direct \CP\ violation
in this decay mode. Unique measurements of \CP\ asymmetries in $\Lambda^0_b$ decays 
are also presented. In the \BsMix\ mixing sector, tensions with SM predictions are now softened by latest updates 
of \phis\ and \DGs\ bounds. From the large data set collected, we also gain a more precise 
measurement of the \BsJpsiPhi\ branching ratio and investigation of dependences of the fragmentation fraction
$f_s/f_d$ on the $B$ transverse momentum. Finally, we obtain the world's best 
measurement of \BsDsDs\ branching ratios to date.
Analyses of the unique ($p\bar{p}$ charge-symmetric) 
and well-understood (10-years expertise) data sample acquired by CDF are still in progress 
and may provide interesting results in the near future.

\end{document}